# Fully Automatic Segmentation of Gross Target Volume and Organs-at-Risk for Radiotherapy Planning of Nasopharyngeal Carcinoma

Mehdi Astaraki[1,2], Simone Bendazzoli[3], Iuliana Toma-Dasu[1,2]

[1]Department of Medical Radiation Physics, Stockholm University, Solna, Sweden
[2]Department of Oncology-Pathology, Karolinska Institutet, Solna, Sweden
[3]Department of Biomedical Engineering and Health Systems, KTH, Huddinge, Sweden
mehdi.astaraki@fysik.su.se

**Abstract**. Target segmentation in CT images of Head&Neck (H&N) region is challenging due to low contrast between adjacent soft tissue. The SegRap 2023 challenge has been focused on benchmarking the segmentation algorithms of Nasopharyngeal Carcinoma (NPC) which would be employed as auto-contouring tools for radiation treatment planning purposes. We propose a fully-automatic framework and develop two models for a) segmentation of 45 Organs at Risk (OARs) and b) two Gross Tumor Volumes (GTVs). To this end, we preprocess the image volumes by harmonizing the intensity distributions and then automatically cropping the volumes around the target regions. The preprocessed volumes were employed to train a standard 3D U-Net model for each task, separately. Our method took second place for each of the tasks in the validation phase of the challenge. The proposed framework is available at https://github.com/Astarakee/segrap2023

**Keywords:** organ segmentation · tumor segmentation · head-neck cancer · autocontouring.

## Introduction

Nasopharyngeal Carcinoma (NPC) is a type of cancer with unique ethnic and geographical distribution. This cancer is highly prevalent in east and southeast Asia which accounts for 120079 newly diagnosed cases and more than 72000 associate deaths only in 2018. NPC, which is considered a subset of Head&Neck (H&N) cancer mostly originates from the epithelium lining of the nasopharynx, a relatively small region located at the back of the nasal cavity [7, 8]. NPC is highly sensitive to ionizing radiation, therefore radiotherapy (RT) is the main treatment modality for non-metastatic disease. Currently, Intensity-Modulated Radiotherapy (IMRT) is the most common technique for NPC treatment. It has been shown that treating with IMRT can reduce the 5-year occurrence rates of locoregional failure for newly diagnosed and localized NPC to 7.4 percent. Recently, there has been great interest in employing proton or carbon-ion RT techniques for treating NPC to further improve the treatment outcomes [2, 10].

MRI, CT, and 18F-fluorodeoxyglucose (18F-FDG)-PET/CT are the most widespread imaging modalities for NPC staging and RT planning. While MRI acquires high-resolution images of soft tissues in H&N regions, CT and MRI images share similar accuracy in detecting cervical lymph node involvement. In addition, 18F-FDG-PET/CT is the standard modality for diagnosing distant metastasis. Acquisition of multimodal imaging data, adaptive radiotherapy techniques, the advent of novel image-guided treatment modalities, and the large number of anatomical structures associated with radiation-induced toxicity led to the accumulation of an overabundance of images to be segmented manually. In fact, the quality of contour delineation of Gross Target Volumes (GTVs) and Oragns-at-Risks (OARs) is a core factor affecting the efficacy and side effects of radiotherapy. Therefore, the development of accurate auto-segmentation methods for the delineation of GTVs and OARs is of great importance to both reduce the delineation time and avoid the inter/intra observer variability issues.

Thanks to the rapid advances of Deep Learning (DL) methods for computer vision applications, DL techniques have been increasingly utilized in medical image segmentation tasks. While numerous DL models have been, specifically, developed for volumetric segmentation of medical images, the conventional encoder-decoder architecture of the U-Net model and its variants remains the most accurate and effective one for a variety of applications. In fact, powerful models such as Attention U-Net, Dense U-Net, U-Net++, and nnU-Net are inspired by the basic U-Net model [11, 12, 5]. For the specific case of NPC, the development of auto-segmentation methods require more attention on the following grounds: (1) More than 40 OARs with various structures and complex shapes need to be carefully delineated. For instance, the delineation of tiny structures such as internal auditory canals, optic nerves, pituitary glands, etc. is demanding even for expert radiologists. (2) The size of the OARs in the H&N region is extremely imbalanced. In other words, while structures such as the brain stem, eyes, trachea, and esophagus occupy thousands of voxels, structures such as the vestibule and lens are presented with only a few voxels. (3) CT images are naturally unable to acquire high-contrast regions from soft tissues. Therefore, discriminating the borders between the adjacent anatomical regions is challenging [1, 4]. To deal with the mentioned difficulties, several methods have been introduced. Tappeiner et al. optimized the patch size of the nnU-Net model based on a class imbalance criterion and then developed a class adaptive Dice loss to tackle the problem of class imbalance [14]. Kawahara et al. proposed a stepwise framework for the segmentation of six OARs. In specific, a U-Net-like network was used as the first step to identify the target regions. Then, the second network was trained on the crop regions resulting from the first step [6]. Zhong et al. followed a human-in-the-loop strategy and managed to improve the segmentation accuracy of 15 OARs. Particularly, they trained a 2D U-Net-like network followed by a subjective evaluation by 2 or 3 oncologists for further retraining of the model [16]. Zhou et al. modified the conventional 3D U-Net model and introduced AnatomyNet. They (a) fed the network with full-size volumes instead of cropping regions, (b) employed 3D squeeze-and-excitation residual blocks in the encoding layers for better feature representation, and (c) combined Dice loss and focal loss as the objective function for model optimization [17]. Tang et al. proposed a two-step pipeline based on candidate region detection, in which local contrast normalization is applied in the segmentation head to improve the segmentation accuracy of 28 OARs [13]. Gao et al. proposed FocusNetv2 to tackle the imbalanced large and small organ segmentation. In fact, their proposed model aimed to resemble how expert radiologists delineate the OARs. Their solution consists of two main components: the segmentation network and the adversarial autoencoder (AAE) for organ shape constraint. The segmentation network first segments all OARs in the first module and localizes the central locations of a series of pre-defined tiny organs in the second module. Then the third module combined the multi-scale features maps with high-resolution images for the task of small OARs segmentation. The AAE constrains the segmentation results with prior shapes of OARs to deal with the problem of boundary ambiguities [4]. In addition to OARs segmentation, in the context of H&N cancer, the head and neck tumor (HECKTOR) segmentation challenge was established to benchmark the developed algorithms for the segmentation of primary gross tumors and nodal gross tumors in PET-CT images [9].

Researchers from the University of Electronic Science and Technology of China in collaboration with the Sichuan cancer center organized the Segmentation of OARs and GTVs of NPC for Radiotherapy Planning (SegRap 2023) challenge aiming at benchmarking the accuracy and robustness of segmentation algorithms.

In this study, we propose a pipeline for the automatic segmentation of OARs and GTVs of NPC in bi-modal CT volumes. By modifying the histogram of each of the CT volumes in advance of the model training, we enhance the contrast of the images. To force the attention of the model toward the target regions, we crop the volumes and discard the background and irrelevant structures. Finally, two models are trained for each of the OARs and GTVs, separately.

## Materials and Methods

### Image data and tasks

The SegRap 2023 challenge incorporates the two tasks. Task 1 concentrates on the segmentation of the following 45 OARs: brain, brainstem, chiasm, left and right cochlea, esophagus, left and right eustachian tubes, left and right eyes, left and right hippocampus, left and right internal auditory canals, larynx, larynx glottic, larynx supraglottic, left and right lens, left and right mandible, left and right mastoid, left and right middle ears, left and right optic nerves, oral cavity, left and right parotid, pharyngeal constrictor muscle, pituitary, spinal cord, left and right submandibular, left and right temporal lobes, thyroid, left and right temporomandibular joints, trachea, left and right tympanic cavities, and left and right vestibular semicircular canals. It should be noted that 9 additional substructures were included in the segmentation labels i.e., a total of 54 labels to deal with the issue of overlapping structures. From the computational perspective, this is a rather special application, since one single model should predict 54 output channels that is in contrast to ordinary applications predicting only a few output channels. Essentially, it means the model requires powerful computational resources to predit such large size tensors. Task 2, on the other hand, focuses on the segmentation of nasopharynx GTV (GTVnx) and lymph nodes GTV (GTVnd).

The dataset consists of CT volumes from 200 patients diagnosed with NPC from which 120 subjects with corresponding pixel-level annotations were released as the training subjects. The developed algorithms should be containerized and submitted on an online platform to evaluate the performance of the trained models on 20 subjects of validation and 60 subjects of test sets. For each subject two co-registered CT volumes were provided; a conventional non-contrast CT and a contrast-enhanced CT. While the in-plane voxel size of the provided scans varies from 0.433 to 1.130 mm, the slice thickness is fixed to 3mm for all subjects. The number of axial slices lies in the range of 98 to 197. In addition, the axial array size of 117 out of 120 training subjects is 1024×1024 pixels while the other 3 subjects were presented with the size of 512×512 pixels. Figure 1 demonstrates an example of a subject with corresponding segmentation labels.

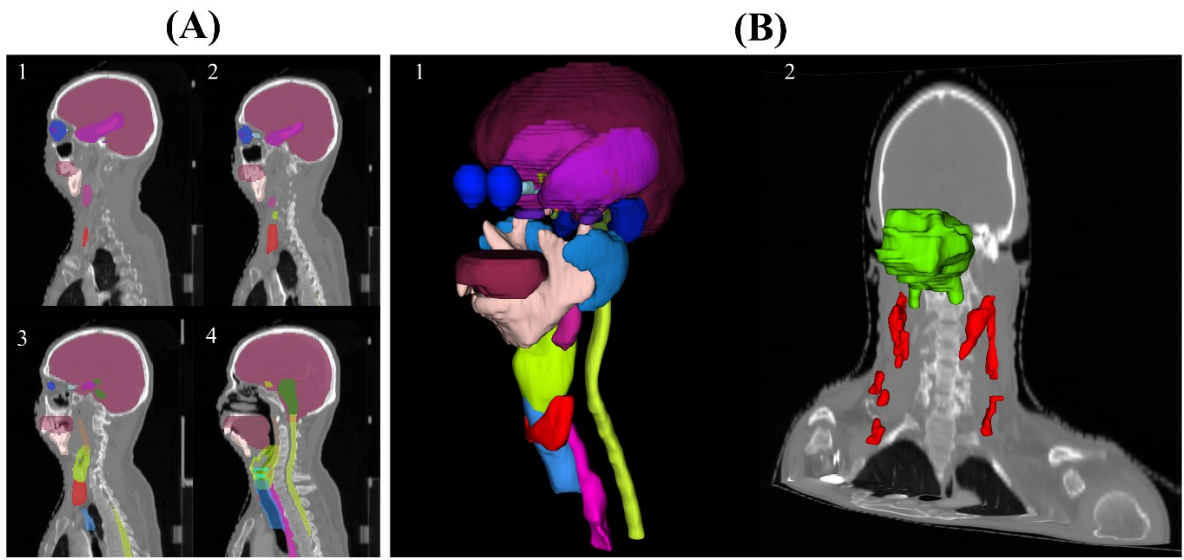

**Fig. 1.** An example of segmentation regions of tasks 1 and 2 in a contrast-enhanced CT image; Subimage (A) shows 4 different 2D sagittal slices of the same subject overlayed with segmentation labels of 45 OARs. Subimage (B1) depicts the volumetric rendering of OARs, and (B2) demonstrates the GTVnx in green and GTVnd in red colors overlayed on the orthogonal views.

**Image preprocessing**

Preparation and preprocessing of the image data are conducted in two steps. First, To enhance the contrast between soft tissues inside the H&N region, Hounsfield values of the CT volumes were clamped into certain ranges depending on the modality and task. In task 1, to better distinguish overlapping OARs from each other the intensity values of contrast-enhanced CTs were clamped into the range of [-400,2000] and [-300,800] for the normal CT volumes. In task 2, to better discriminate the pathological regions from nearby healthy tissues, the Hounsfield values of contrast CTs were clamped into the range of [-1000,1000] and to the range of [-600,600] for the normal CTs. It is worth mentioning that the described intensity values were chosen experimentally after visual examination of several subjects for each task and each modality independently. Figure 2 illustrates the effects of the intensity clamping on the image data.

In the second step, we cropped the CT volumes to discard the background and irrelevant anatomical structures. In fact, the provided high-resolution CT images (1024×1014) contain anatomical regions from the top head to the middle of the thorax region. However, only a narrow region from such wide arrays incorporates the H&N-related structures. Therefore, to avoid computational complexity and speed up the processing time, the following steps were applied to cropping the volumes into the target regions. In practice, the TotalSegmentor [15] model was employed to segment the 'body' structures. From its outputs, we picked the 'body-extremities' segmentation mask which contains the H&N regions as well as other lower-body limbs. In order to exclude the irrelevant limbs, a connected component (CC) analysis was conducted, and a bounding box was fitted to the largest CC which includes only the head-neck regions. Since some OARs such as the trachea, esophagus, spinal cord, etc. are extended toward the thorax region, all the axial slices were included in the bounding box as well. Finally, a margin of 15 pixels was considered to widen the detected bounding box in order to compensate for the potential false negatives resulting from the TotalSegmentor. Figure 3 visualizes the result of the cropping procedure.

**Segmentation models**

By employing the training data set, we trained two independent models for tasks 1 and 2. While the inputs of the two models are bi-modal CT volumes, the segmentation masks are different, i.e., they contain 54 labels for OARs segmentation and 2 labels for GTVs. The nnU-Net V1 framework was adopted as the segmentation pipeline. The two models were trained with a 5-fold cross-validation fashion. Table 1 summarizes the details of the training protocols.

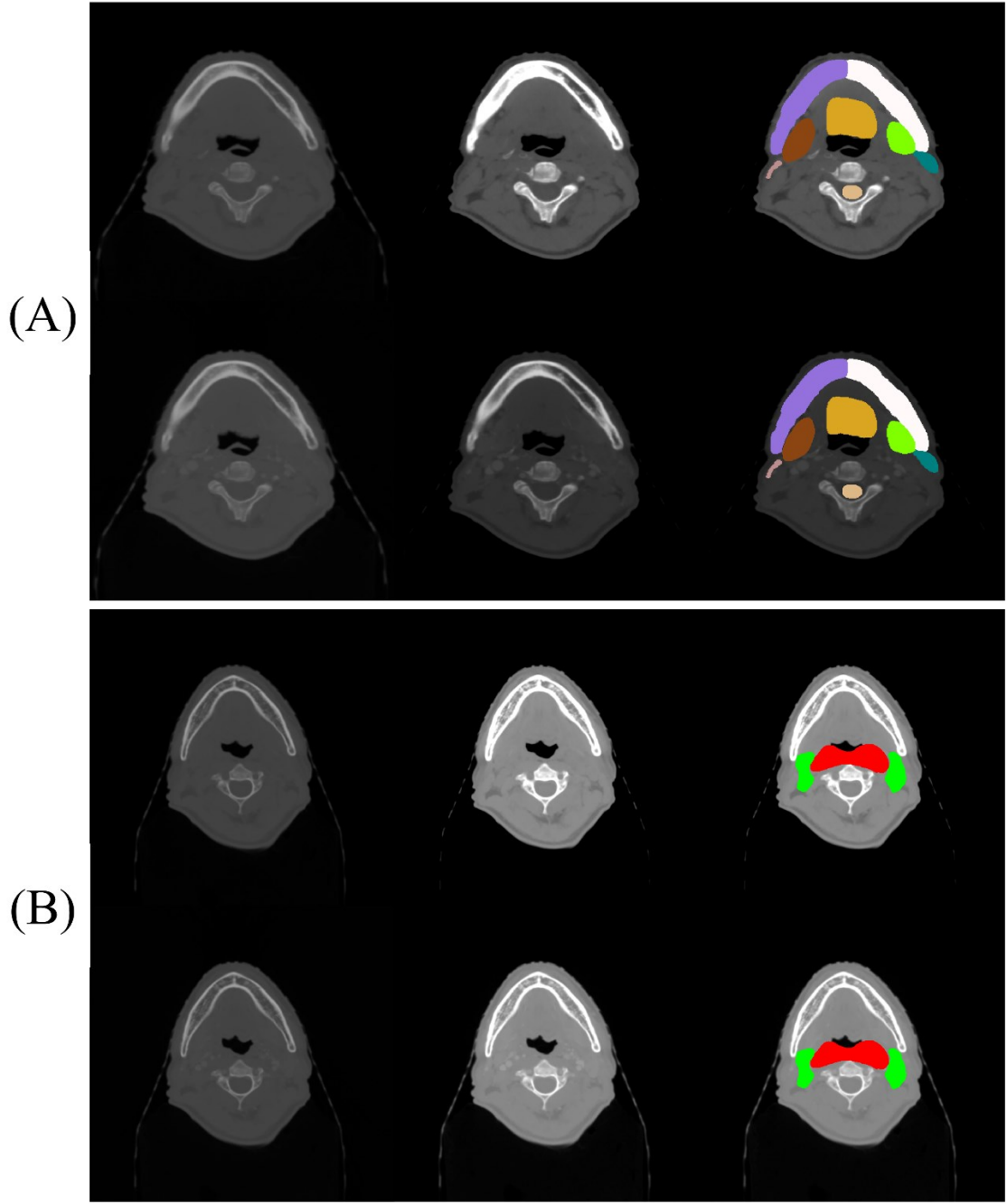

**Fig. 2.** Intensity windowing for contrast enhancement. Subimage (A) illustrates examples regarding task 1 and subimage (B) visualizes examples related to task 2. For each of the subimages, the first row depicts the ordinary CTs and the second row shows the enhanced CTs. From left to right: first images show the original slices, middle images present the contrast-enhanced images and third ones illustrate the overlayed segmentation labels.

In practice, the OARs model was trained with full-resolution images. However, in the inference phase, we, first, cropped the volumes as mentioned in the preprocessing section and then predict the segmentation labels over the cropped images. On the other hand, the GTVs model was trained with cropped volumes instead of full-resolution images. In our experiments, the default value of 1000 epochs was not enough to converge the learning curves of task 1. In other words, the complexity of learning 54 structures simultaneously, inevitably, requires more training iterations. Therefore, the number of training epochs was increased to 2500 for each fold. On the other hand, despite the applied data augmentation, 1000 epochs were too much for task 2 as it was prone to overfitting the model. Hence, GTVs model was trained for 600 epochs. Figure 4 shows a sample learning curve of each model.

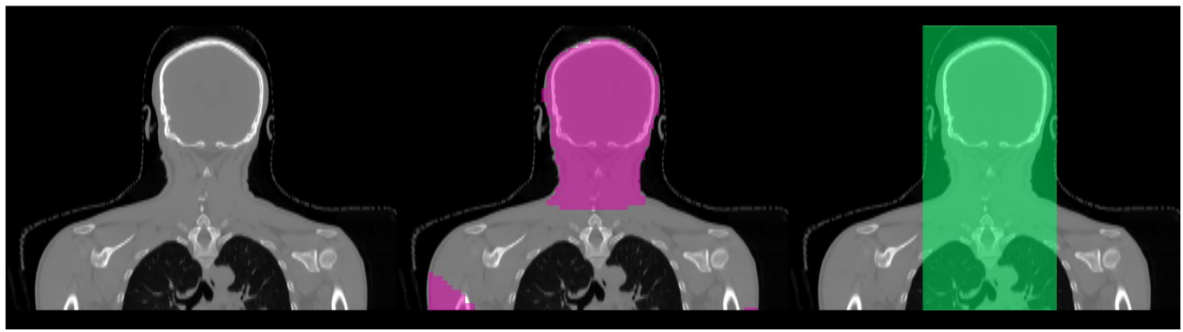

**Fig. 3.** CT volumes were cropped around the target H&N region. From left to right: the first image shows a coronal view of a subject; the result of the TotalSegmentor is overlayed on the middle image; and the final bounding box is highlighted in green color on the last image.

Table 1. Hyperparameters of the segmentation models

| Parameters | OARs model | GTVs model |
|---|---|---|
| Trainer class | nnUnetTrainerV2 | nnUnetTrainerV2 |
| Objective function | Dice + BCE | Dice + BCE |
| Optimizer | SGD | SGD |
| Augmentation | True except for the flipping | True |
| Patch size | 64×192×160 | 80×192×128 |
| # of feature maps in the base layer | 32 | 32 |
| # of pooling per axis | [4,5,5] | [4,5,5] |
| # of epochs | 2500 | 700 |
| # of training batches per epoch | 250 | 250 |
| # of validation batches per epoch | 50 | 50 |
| Initial learning rate | 0.01 | 0.01 |
| Batch size | 2 | 2 |
| # of folds | 5 | 5 |

## Experiments and Results

The segmentation accuracy of the OARs model and GTVs model are reported in this section.

The OARs model was developed by cross-validating the training dataset to segment 54 anatomical structures simultaneously. As a postprocessing step, 9 of the overlapping areas are merged to form the expected 45 regions. The performance of the developed model was quantified by employing standard segmentation metrics including Dice, Precision, and Recall. However, to avoid confusion for the readers, we categorized the quantitative values into the following ranges: x≥0.90, 0.90>x≥0.80, 0.80>x, and reported only the Dice metrics in Figure 2AC. Figure 2D shows the results of the cross-validated GTVs model in terms of the Dice metric for the training dataset.

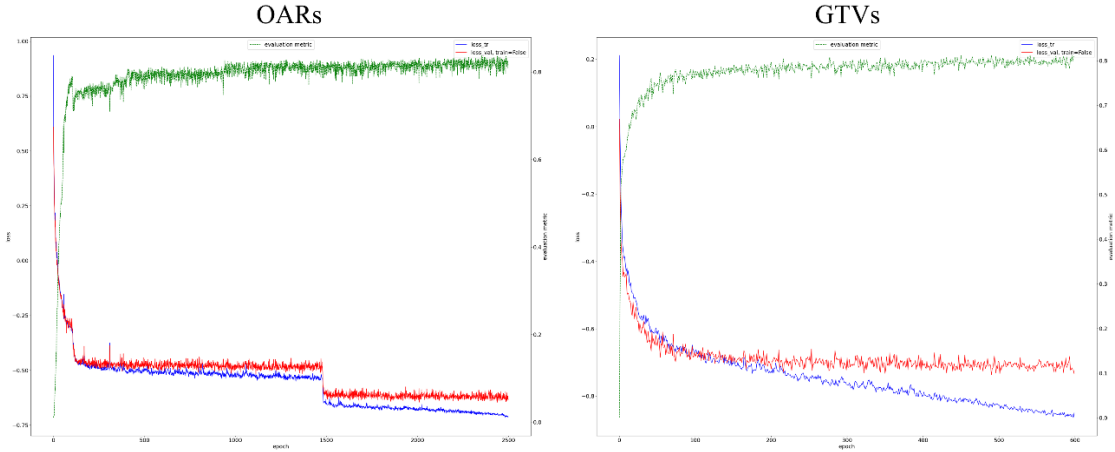

**Fig. 4.** Example of learning curves of the trained models.

In addition, the average Dice metrics over all the 54 OARs of task 1 and 2 GTVs of task 2 are reported in Table 2.

**Table 2.** The overall Dice metrics calculated over the cross-validated training set.

| Model | Dice Similarity Metric | |
|---|---|---|
| | Mean±std | Median |
| GTVs | 0.758±0.080 | 0.776 |
| OARs | 0.851±0.061 | 0.845 |

The developed models were dockerized and submitted to the challenge online platform to predict the segmentation labels of the validation set. In this context, it should be noted that the challenge organizers set a memory limit of a maximum of 32 GB to run the dockerized algorithm. However, ensembling the softmax resulted from 5 folds to predict the 54 segmentation labels requires more than 50 GB of system memory. Accordingly, in order to meet the memory restriction, we trained another model for task 1 with the same parameters mentioned in Table 1 over all the training datasets i.e. all 120 subjects were used altogether for the training procedure. Additionally, in the inference phase, the interpolation order was reduced and test time augmentation was disabled as well. By posing the described modifications, the models were submitted and successfully executed over the validation subjects. Table 3 summarizes the reported quantification metrics by the challenge organizers.

**Table 3.** The overall quantified metrics over the validation set.

| Model | Mean±std | |
|---|---|---|
| | Dice | Normalized Surface Dice |
| GTVs | 0.790±0.089 | 0.739±0.119 |
| OARs | 0.887±0.076 | 0.905±0.087 |

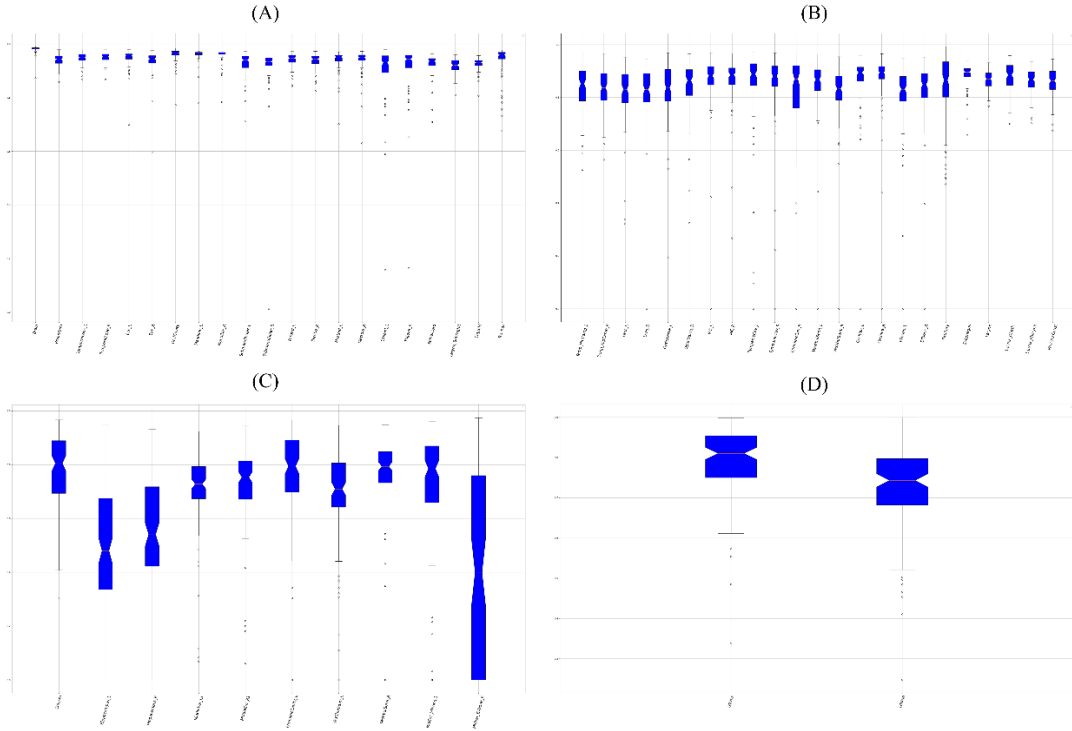

**Fig. 5.** Boxplot visualization of the quantified Dice metrics. A) Structures in OARs model with Dice>=0.90; B) Structures in OARs model with 0.80<=Dice<0.90; C) Structures in OARs model with Dice<80; and D) Dice metrics of GTVs model.

## Discussion and Conclusion

Analyzing the medical images to characterize the attributes of tumors and predict the treatment outcomes has shown great potential to personalize and optimize cancer care, especially in the context of H&N cancers [9, 3]. In the case of NPC, CT is the standard imaging modality for diagnosing and planning interventions, in particular radiation treatment. Yet, manual or semi-automatic delineation of GTVs and OARs in radiotherapy planning is a subjective, error-prone, and tedious task. Therefore, the fully automatic delineation of tumor(s) and healthy organs with accurate 3D segmentation models can dramatically speed up this process and further provide reproducible definitions of GTVs and OARs for both planning systems and imaging biomarkers studies. The Segmentation of OARs and GTVs of NPC for Radiotherapy Planning (SegRap 2023) challenge aims at identifying the best auto contouring methods by evaluating the segmentation accuracy of the submitted algorithms developed over bi-modal scans including CTs and contrast-enhanced CTs.

In this study, we proposed a pipeline for the automatic segmentation of 45 OARs and 2 GTVs which includes 3 steps: volume preprocessing, target segmentation, and post-processing. In the first step, we enhanced the contrast of raw volumes by clamping their intensity distributions into fixed ranges. This strategy not only enhances the contrast between the soft tissues but also harmonizes the histograms of the input images. In other words, while the histogram of acquired images varies significantly between the subjects, applying an intensity windowing strategy will bring all the variations into a certain range. This harmonization step will further help the channel-wise Z-scoring intensity normalization to rescale all the input voxel values into a fixed range which facilitates the training procedure. In addition, we cropped the volumes around the target regions and discarded the background and irrelevant structures. This strategy is beneficial for the learning process because the model can mainly focus on the candidate regions and not be confused by other similar but irrelevant anatomical structures. Another advantage of cropping the volumes is to save computational time and resources specifically during the inference phase. In specific, the size of the cropped arrays is usually smaller than half of the size of the original arrays toward the width and height axes. As a result, the prediction time is decreased by up to 8 times and the required system memory is divided by half.

The nnU-Net pipeline was employed to develop two different segmentation models for Task 1 and Task 2. The superiority of the nnU-Net was already validated in quite many different applications [5]. In task 1, out of 54 OARs, 21 structures were segmented with a mean Dice metric larger than 0.90, 23 structures were segmented with a mean Dice in the range of 0.80 to 0.90, and 10 structures ended up with a mean Dice less than 0.80 from which only 3 OARs were segmented poorly (less than 0.60). Considering the large variations in the size of the OARs, the model could successfully segment all the structures and resulted in an average Dice metrics of 0.851 on the training set and 0.887 on the validation set. In the lack of molecular imaging, FDG-PET modality, detection, and delineation of GTVs will become challenging and heavily depends on the anatomical changes and deformations. Nevertheless, the developed model for task 2 learned such deviations and yielded rather robust results, i.e., the average Dice metric is more than 0.75 for both training and validation sets. In the postprocessing step, the coordinates of the cropped regions were employed to reconstruct the final full-resolution segmentation masks.

Finally, despite the promising results achieved by the proposed pipeline for both OARs and GTVs segmentation, there are some limitations that will be investigated in our future studies. Specifically, a weighting strategy can be applied to the objective function in order to compensate for the imbalanced size of OARs and further improve the segmentation accuracy of tiny structures. Furthermore, the potential of simultaneous segmentation of GTVs and OARs in a single model can be investigated to take into account the mass effects.

In conclusion, we have proposed a segmentation pipeline for auto-contouring the OARs and GTVs of NPC for radiation therapy planning. The performance of this framework was validated which verified the accuracy and robustness of the model.

## Acknowledgment

This study was supported by the Cancer Research Funds of Radiumhemmet. The Swedish Cancer Society and The Swedish Research Council (2020-04618) is gratefully acknowledged. We would like to appreciate Stockholm Medical Image Laboratory and Education (SMILE) for giving us access to their Nvidia DGX-1 cluster.